\documentclass[twocolumn,showkeys,aps,prb,showpacs]{revtex4-1}
\usepackage{graphicx}
\usepackage[CJKbookmarks,dvipdfm,colorlinks,linkcolor=blue,citecolor=blue]{hyperref}

\begin{document}

\title{Tuning pure out-of-plane piezoelectric effect of  penta-graphene: a first-principle study}

\author{San-Dong Guo}
\affiliation{School of Electronic Engineering, Xi'an University of Posts and Telecommunications, Xi'an 710121, China}
\begin{abstract}
For  two-dimensional (2D) materials, a  pure large out-of-plane piezoelectric response,  compatible with the nowadays bottom/top gate technologies, is
highly desired. In this work, the  piezoelectric properties  of  penta-graphene (CCC) monolayer   are studied  with pure out-of-plane piezoelectric effect by density functional theory (DFT).  However,  the $d_{36}$ is very small, and only -0.065 pm/V.
Two strategies are proposed to enhance  piezoelectric properties  of  CCC  monolayer. Firstly, both  biaxial and uniaxial  strains are applied,
but the enhancement is very small, and at -2\%  biaxial (-4\% uniaxial) strain, the $d_{36}$ is  increased only by 3.1\% (13.9\%). Secondly, a Janus penta-monolayer (CCB) is  constructed by replacing the top C (B) atomic layer in monolayer CCC [pentagonal $\mathrm{CB_2}$ monolayer (CBB)] with B (C) atoms,  which shows  dynamic and mechanical stability. Fortunately,  the pure out-of-plane piezoelectric effect of CCB monolayer still holds, and exhibits a band gap.  The calculated  $d_{31}$ and $d_{32}$  are   -0.505 pm/V  and 0.273 pm/V, respectively, which are very larger than  $d_{36}$ of CCC monolayer. The out-of-plane piezoelectricity $d_{31}$ of CCB monolayer is obviously higher compared with many other 2D known materials.
Moreover, its room-temperature electronic mobility along y direction  is as high as 8865.23 $\mathrm{cm^2V^{-1}s^{-1}}$.
Our works provide  a new way to achieve pure out-of-plane piezoelectric effect, which is highly desirable for ultrathin piezoelectric devices.

\end{abstract}

\pacs{71.20.-b, 77.65.-j, 72.15.Jf, 78.67.-n ~~~~~~~~~~~~~~~~~~~~~~~~~~~~~~~~~~~Email:sandongyuwang@163.com}

\maketitle

\section{Introduction}
 Beyond graphene, a large amount of 2D carbon allotropes have  been  investigated. The  CCC monolayer of them, composed entirely of carbon pentagons, could be realized experimentally  with good thermodynamic stability\cite{w1}. Inspired from CCC monolayer, other
 pentagon-based 2D materials have been theoretically predicted  or experimentally
synthesized, such as
penta-$\mathrm{CB_2}$\cite{w2},  penta-$\mathrm{SnX_2}$ (X=S, Se, or Te)\cite{w3} and penta-$\mathrm{AlN_2}$\cite{w4}.
The mechanical behavior of monolayer CCC under multi-axial loading has been studied by DFT calculations, and  the structure has lower ultimate tensile strength  compared to graphene\cite{w5}.   The  phonon transport properties of CCC monolayer  have been widely investigated theoretically, and the intrinsic lattice thermal conductivity is significantly reduced as compared to that of graphene\cite{w6,w6-1,w6-2}.  The magnetic moments can be induced  by an isolated hydrogen atom absorbed on  CCC monolayer, which  changes CCC monolayer from a semiconductor to half-metallic\cite{w7}.
However, to the best of our knowledge, piezoelectric properties of CCC monolayer haven't  been reported.

In fact,  the piezoelectricity in 2D materias has
attracted growing interest\cite{q4} because of  potential nanoscale piezoelectric
applications, like sensors, actuators and energy sources.   For 2D materials, the reduction in dimensionality  makes their inversion symmetry
disappear,  which  allows them to become piezoelectric. Experimentally,  the monolayer $\mathrm{MoS_2}$  with  the
2H phase has been  proved to be  piezoelectric  ($e_{11}$=2.9$\times$$10^{-10}$ C/m)\cite{q5,q6}, and  the existence of vertical dipoles in the Janus MoSSe monolayer  has also  been  observed,  showing an intrinsic vertical piezoelectric response\cite{q8}.
In theory, lots of studies on  piezoelectric properties related with 2D materials  have been reported\cite{q7,q9,q10,q11,q12}, like  transition metal dichalchogenides (TMD), group IIA and IIB metal oxides, group III-V semiconductors, group-V binary semiconductors and Janus TMD.
It is surprising that the giant piezoelectricities  in monolayer SnSe,
SnS, GeSe and GeS  have been predicted by the first-principle calculatios\cite{q10},  as high as  75-251 pm/V  along the  the armchair  direction.
Some 2D materials only exhibit  an in-plane piezoelectricity, like TMD monolayers\cite{q9}, and an additional out-of-plane piezoelectricity has been observed in many  2D  materials, like Janus TMD\cite{q7}. A  pure large out-of-plane piezoelectric response is highly desired, which is compatible with the nowadays bottom/top gate technologies. However, the  pure  out-of-plane piezoelectric response in 2D materials is rarely reported.
\begin{figure}
  \includegraphics[width=8.0cm]{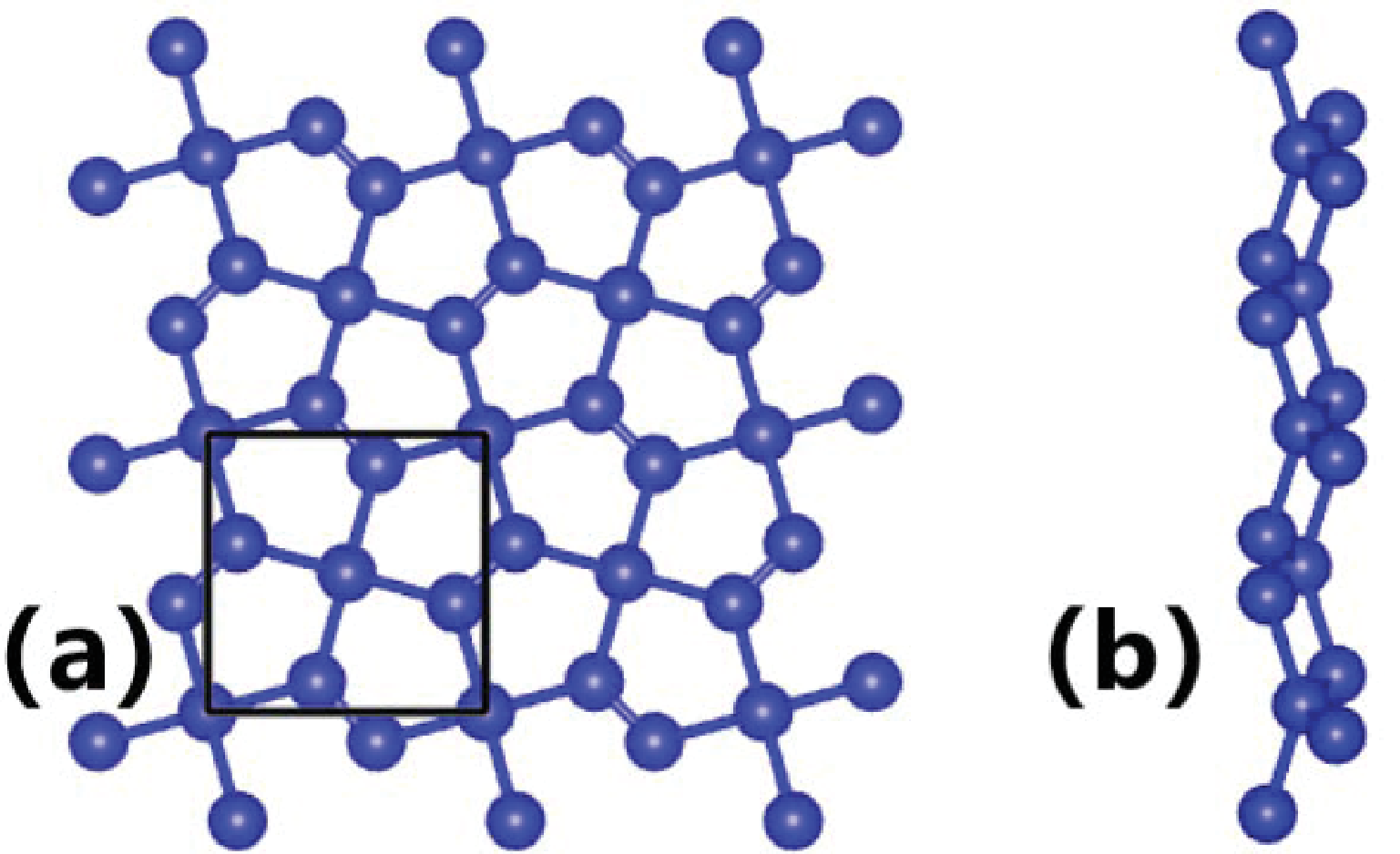}
  \caption{(Color online)The crystal structure of penta-graphene: top view (a) and side view (b), and the primitive cell is
   are marked by black line.}\label{t0}
\end{figure}

In this work,  the piezoelectric properties of CCC monolayer  are reported by using density functional perturbation theory (DFPT)\cite{pv6} with generalized gradient approximation (GGA).  Only  out-of-plane $d_{36}$  exists for CCC monolayer, but it is very small.  Both  biaxial and uniaxial  strains are used to tune its piezoelectric properties, but the improvement is very small. Janus monolayer can be built from symmetric sandwich structure, like MoSSe synthesized by replacing the top S atomic layer in  $\mathrm{MoS_2}$ with Se atoms\cite{q8}. Inspiring from  MoSSe monolayer,  a Janus CCB monolayer with dynamic and mechanical stability  is  constructed, and the special symmetry leads to only  out-of-plane piezoelectric effect.  Compared with  $d_{36}$ of CCC monolayer, the predicted $d_{31}$ and $d_{32}$ obviously are  improved, the  $d_{31}$ of which  is  higher compared with other many 2D known materials.
Another significant advantage for CCB monolayer is a very high  room-temperature electronic mobility (8865.23 $\mathrm{cm^2V^{-1}s^{-1}}$) along y direction.
Therefore, our works give an experimental proposal  to achieve pure out-of-plane piezoelectricity in 2D materials, and pave a way for  designing piezoelectric devices compatible with the nowadays bottom/top gate technologies.

The rest of the paper is organized as follows. In the next
section, we shall give our computational details and methods  about piezoelectric coefficients. In the third and fourth sections, we shall present piezoelectric properties of monolayer CCC and CCB. Finally, we shall give our  conclusions in the fifth section.

\begin{figure}
  \includegraphics[width=8cm]{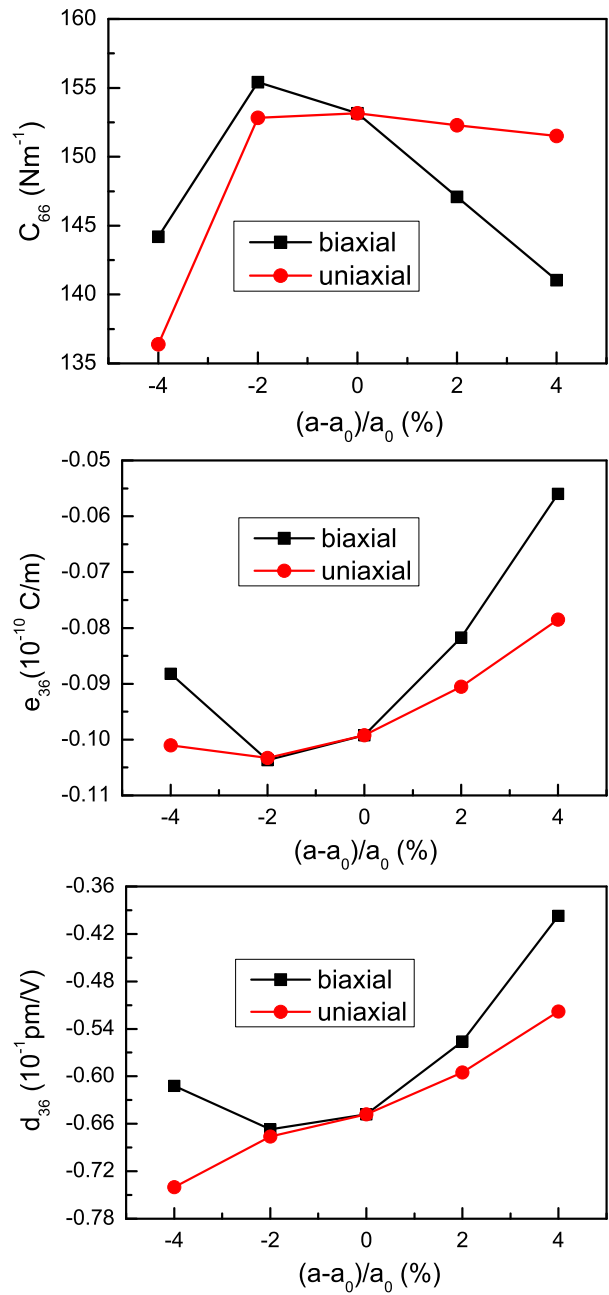}
  \caption{(Color online) For monolayer CCC, the elastic constants $C_{66}$,  piezoelectric coefficients $e_{36}$ and $d_{36}$   with the application of  biaxial and uniaxial  strains.}\label{sb}
\end{figure}

\begin{table*}
\centering \caption{For CCC, CCB and CBB monolayers, the lattice constants $a_0/b_0$ ($\mathrm{{\AA}}$) and $\gamma$,  the GGA  gaps  (eV), and the elastic constants $C_{ij}$ ($\mathrm{Nm^{-1}}$).}\label{tab0}
  \begin{tabular*}{0.96\textwidth}{@{\extracolsep{\fill}}cccccccc}
  \hline\hline
Name& $a_0/b_0$ &  $\gamma$& Gap&$C_{11}$&  $C_{22}$&  $C_{12}$&  $C_{66}$     \\\hline
CCC &3.639&90&2.20& 271.84 & 271.84&-20.86 &153.15\\\hline
CCB& 3.771& 92.107& 0.66&142.36 &¡¡ 193.07& 34.40& 61.71\\\hline
CBB&3.933&90 & 1.41&86.78 &86.78 &87.86 &89.20\\\hline\hline
\end{tabular*}
\end{table*}

\section{Computational detail}
 Within the framework of DFT\cite{1}, our calculations are performed by using the  VASP package\cite{pv1,pv2,pv3}.
 The projected augmented wave
(PAW) method with a kinetic cutoff energy of 500 eV is adopted, and  we use the popular GGA of Perdew, Burke and  Ernzerhof  (GGA-PBE)\cite{pbe} as the exchange-correlation potential. For all studied monolayers, a vacuum spacing of
 more than 16 $\mathrm{{\AA}}$ along the z direction is included to avoid interactions
between two neighboring images.
The total energy  convergence criterion is set
to $10^{-8}$ eV, and  the Hellmann-Feynman forces  on each atom are less than 0.0001 $\mathrm{eV.{\AA}^{-1}}$.
The coefficients of the elastic stiffness tensor  $C_{ij}$ are calculated by using the finite difference method (FDM), and the piezoelectric stress coefficients $e_{ij}$ are calculated by  DFPT method\cite{pv6}.
 Within FDM and DFPT, the electronic and ionic contribution to
the elastic and  piezoelectric stress coefficient can be attained  directly from VASP code.
For $C_{ij}$ and $e_{ij}$, the Brillouin zone sampling
is done using a Monkhorst-Pack mesh of 23$\times$23$\times$1 for CCC monolayer and CBB monolayer, and  18$\times$19$\times$1 for CCB monolayer.
The 2D elastic coefficients $C^{2D}_{ij}$
 and   piezoelectric stress coefficients $e^{2D}_{ij}$
have been renormalized by the the length of unit cell along z direction ($Lz$):  $C^{2D}_{ij}$=$Lz$$C^{3D}_{ij}$ and $e^{2D}_{ij}$=$Lz$$e^{3D}_{ij}$.

\begin{table}
\centering \caption{For CCC, CCB and CBB monolayers, piezoelectric coefficients $e_{ij}$ and $d_{ij}$ , along with out-of-plane piezoelectric coefficients of some typical 2D
 materials, like SbTeI\cite{q7-2-1}, BiTeI\cite{q7-2-1}, MoSSe\cite{q7-2}  and MoSTe\cite{q7-1}. The unit is $10^{-10}$C/m  for $e_{ij}$ (pm/V for $d_{ij}$). }\label{tab-y}
  \begin{tabular*}{0.48\textwidth}{@{\extracolsep{\fill}}ccccccc}
  \hline\hline
Name & $e_{31}$ & $d_{31}$& $e_{32}$&$d_{32}$&$e_{36}$&$ d_{36}$\\\hline\hline
CCC&    & &&&-0.099& -0.065\\\hline
CCB&  -0.624&  -0.505& 0.353&0.273\\\hline
CBB&      & &&&-0.372&-0.418                            \\\hline\hline
SbTeI&-0.13&-0.37&-0.13&-0.37\\\hline
BiTeI&  -0.23 & -0.66 &  -0.23 & -0.66                              \\\hline
$\mathrm{MoSSe}$&0.42 & 0.29&0.42 & 0.29\\\hline
$\mathrm{MoSTe}$&0.5 &  0.4 &0.5 &  0.4 \\\hline\hline
\end{tabular*}
\end{table}

\begin{figure}
  \includegraphics[width=7.0cm]{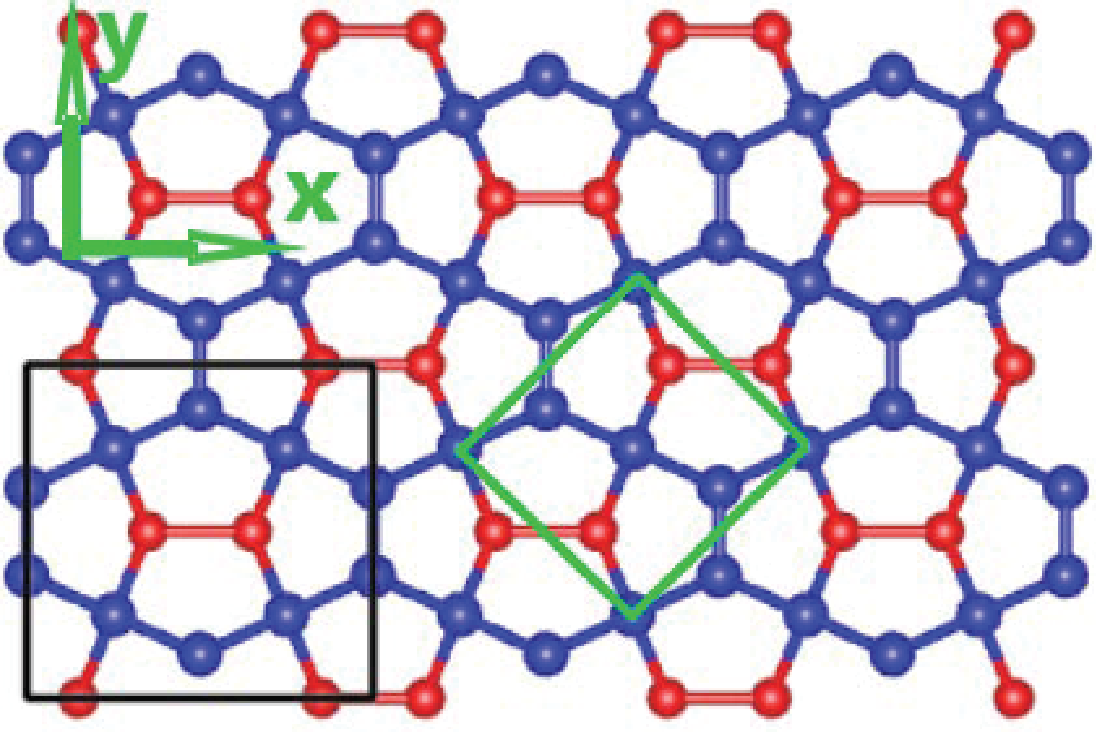}
  \caption{(Color online) The top view of  crystal structure of CCB, and the primitive cell and rectangle supercell
   are marked by green and  black lines, respectively.}\label{t0-1}
\end{figure}

\section{Piezoelectric properties of CCC monolayer}
The strain or stress can induce a change of polarization in noncentrosymmetric crystals, described by the third-rank piezoelectric stress tensors  $e_{ijk}$ and strain tensor $d_{ijk}$. They from the sum of ionic
and electronic contributions  are defined as:
 \begin{equation}\label{pe0}
      e_{ijk}=\frac{\partial P_i}{\partial \varepsilon_{jk}}=e_{ijk}^{elc}+e_{ijk}^{ion}
 \end{equation}
and
 \begin{equation}\label{pe0-1}
   d_{ijk}=\frac{\partial P_i}{\partial \sigma_{jk}}=d_{ijk}^{elc}+d_{ijk}^{ion}
 \end{equation}
Where $P_i$, $\varepsilon_{jk}$ and $\sigma_{jk}$ are polarization vector, strain and stress, respectively.
By employing Voigt notation, for 2D materials with only considering in-plane strain components\cite{q7,q9,q10,q11,q12},
 the  $d_{ij}$ can be derived  using the relation:
 \begin{equation}\label{pe}
  \left(
    \begin{array}{ccc}
      e_{11} & e_{12} & e_{16} \\
     e_{21} & e_{22} & e_{26} \\
      e_{31} & e_{32} & e_{36} \\
    \end{array}
  \right)
  =
  \left(
    \begin{array}{ccc}
      d_{11} & d_{12} & d_{16} \\
      d_{21} & d_{22} & d_{26} \\
      d_{31} & d_{32} & d_{36} \\
    \end{array}
  \right)
    \left(
    \begin{array}{ccc}
      C_{11} & C_{12} & C_{16} \\
     C_{21} & C_{22} &C_{26} \\
      C_{61} & C_{62} & C_{66} \\
    \end{array}
  \right)
   \end{equation}
Where the $C_{ij}$ is elastic tensor, which can be calculated by  FDM,  and the  $e_{ij}$  can be attained by DFPT.
The CCC monolayer has $P\bar{4}2_1m$ symmetry (space
group number 113), and the corresponding point group $\bar{4}2m$ makes \autoref{pe} become:
 \begin{equation}\label{pe1}
  \left(
    \begin{array}{ccc}
     0 & 0 & 0 \\
    0 &0 & 0 \\
      0 & 0 & e_{36} \\
    \end{array}
  \right)
  =
  \left(
    \begin{array}{ccc}
        0 & 0 & 0 \\
    0 &0 & 0 \\
      0 & 0 & d_{36} \\
    \end{array}
  \right)
    \left(
    \begin{array}{ccc}
      C_{11} & C_{12} &0 \\
     C_{12} & C_{11} &0 \\
     0 & 0 & C_{66} \\
    \end{array}
  \right)
   \end{equation}
Here, the  $d_{36}$ is derived by  \autoref{pe1}:
\begin{equation}\label{pe2-7}
    d_{36}=\frac{e_{36}}{C_{66}}
\end{equation}
It is clearly seen that only out-of-plane piezoelectric effect can be observed.

The penta-graphene can be considered as a sandwich structure, with
the 4-coordinated C atoms  sandwiched between the 3-coordinated
atoms, and the schematic structure is plotted in \autoref{t0}.
Firstly,  the lattice constants of CCC are optimized ($a$=$b$=3.639 $\mathrm{\AA}$), which is very close to previous theoretical values\cite{w1}. The elastic stiffness coefficients $C_{ij}$ and piezoelectric stress tensors $e_{ij}$ are calculated, and then $d_{ij}$  can be attained.
By using FDM, we obtain $C_{11}$=271.84 $\mathrm{Nm^{-1}}$  $C_{12}$=-20.86 $\mathrm{Nm^{-1}}$ and  $C_{66}$=153.15 $\mathrm{Nm^{-1}}$.
The calculated $C_{ij}$ agree well with previous ones\cite{w1}. The  $C_{12}$ is negative, which means a negative Poisson's ratio.
These related  data are listed \autoref{tab0}.  For 2D materials, a pure out-of-plane piezoelectric response,  compatible with the nowadays bottom/top gate technologies, is highly desired. For CCC monolayer, only out-of-plane piezoelectric response exits, but the predicted piezoelectric coefficient $d_{36}$ is very small, and the calculated value: $e_{36}$=-0.099$\times$$10^{-10}$ C/m    and $d_{36}$=-0.065 pm/V. Some strategies should be applied to improve  piezoelectric effect  of CCC monolayer, and that only out-of-plane piezoelectric response holds.

It has been proved that strain can effectively improve the  piezoelectric response of 2D materials, such as $\mathrm{MoS_2}$\cite{r1}, AsP\cite{r2}, SnSe\cite{r2} and Janus TMD monolayers\cite{r3}.  For example, the $d_{22}$  of SnSe monolayer at -3.5\% strain along the armchair  direction is up to 628.8 pm/V  from unstrained 175.3 pm/V\cite{r2}. Here, the small both  biaxial and uniaxial  strain (-4\% to 4\%) effects on  piezoelectric properties of monolayer CCC are investigated.
The  elastic constants $C_{66}$, piezoelectric coefficients $e_{36}$  and  $d_{36}$  as a function of strain
are plotted in \autoref{sb}.  With both   biaxial and uniaxial  strains changing from -4\% to 4\%, both $C_{66}$ and $e_{36}$ show
a non-monotonic behavior. It is found that, according to \autoref{pe2-7}, the compressive strain  is in favour of improving  piezoelectric response due to reducing $C_{66}$ and improving $e_{36}$ (absolute value). However, the enhancement is very small, and at
-2\%  biaxial (-4\% uniaxial) strain, the $d_{36}$ becomes   -0.067 (-0.074
) pm/V   from unstrained -0.065 pm/V, increased only by 3.1\% (13.9\%).

\section{Piezoelectric properties of CCB monolayer}
A pentagonal CBB monolayer has also been predicted\cite{w2},  which can also be viewed as a
B-C-B sandwich trilayer. In other words, the C atoms of  the first and third layers of penta-graphene are replaced by B atoms.
The $P\bar{4}2_1m$ symmetry still holds for CBB monolayer, and our optimized lattice constants $a$=$b$=3.933 $\mathrm{\AA}$, which agrees well with previous ones\cite{w2}. The calculated piezoelectric coefficients are  $e_{36}$=-0.372$\times$$10^{-10}$ C/m    and $d_{36}$=-0.418 pm/V, which are very larger than ones of CCC monolayer. However, our calculated elastic stiffness coefficients ($C_{11}$=86.78 $\mathrm{Nm^{-1}}$,  $C_{12}$=87.86 $\mathrm{Nm^{-1}}$ and  $C_{66}$=89.20 $\mathrm{Nm^{-1}}$) violate the  Born  criteria of mechanical stability\cite{ela1,ela}:
\begin{equation}\label{ela}
C_{11}C_{22}-C_{12}^2>0~ and ~C_{66}>0
\end{equation}
Experimentally, Janus TMD monolayer MoSSe with sandwiched S-Mo-Se structure has been synthesized by replacing the top S atomic layer in  $\mathrm{MoS_2}$ with Se atoms\cite{q8}.  Thus, it is possible to build  CCB monolayer by replacing the top C (B) atomic layer in monolayer CCC (CBB) with B (C) atoms. The schematic structure of CCB monolayer is shown in  \autoref{t0-1}.  The symmetry of CCB monolayer reduces to $Cmm2$ (space
group number 35), and the corresponding point group is $mm2$.  The \autoref{pe} changes into:
\begin{equation}\label{pe1-ccb}
  \left(
    \begin{array}{ccc}
     0 & 0 & 0 \\
    0 &0 & 0 \\
     e_{31} &e_{32} &0  \\
    \end{array}
  \right)
  =
  \left(
    \begin{array}{ccc}
        0 & 0 & 0 \\
    0 &0 & 0 \\
      d_{31} &d_{32} & 0 \\
    \end{array}
  \right)
    \left(
    \begin{array}{ccc}
      C_{11} & C_{12} &0 \\
     C_{12} & C_{22} &0 \\
     0 & 0 & C_{66} \\
    \end{array}
  \right)
   \end{equation}
Here, the  $d_{31}$ and $d_{32}$ are derived by  \autoref{pe1-ccb}:
\begin{equation}\label{pe2-7-ccb}
    d_{31}=\frac{C_{22}e_{31}-C_{12}e_{32}}{C_{11}C_{22}-C_{12}^2}
\end{equation}
\begin{equation}\label{pe2-7-ccb}
    d_{32}=\frac{C_{11}e_{32}-C_{12}e_{31}}{C_{11}C_{22}-C_{12}^2}
\end{equation}
Fortunately, the CCB monolayer also possesses only out-of-plane piezoelectric effect.
These also   imply  that the piezoelectric effects of  CCB monolayer can be induced
  with a uniaxial strain being applied along x or/and y direction, which is different monolayer CCC and CBB with shear strain.

\begin{figure}
  \includegraphics[width=7cm]{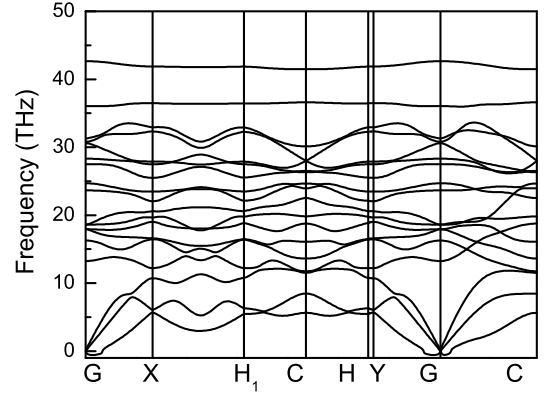}
\caption{The phonon band dispersion  of  CCB monolayer. }\label{t1-1}
\end{figure}
\begin{figure*}
  \includegraphics[width=12cm]{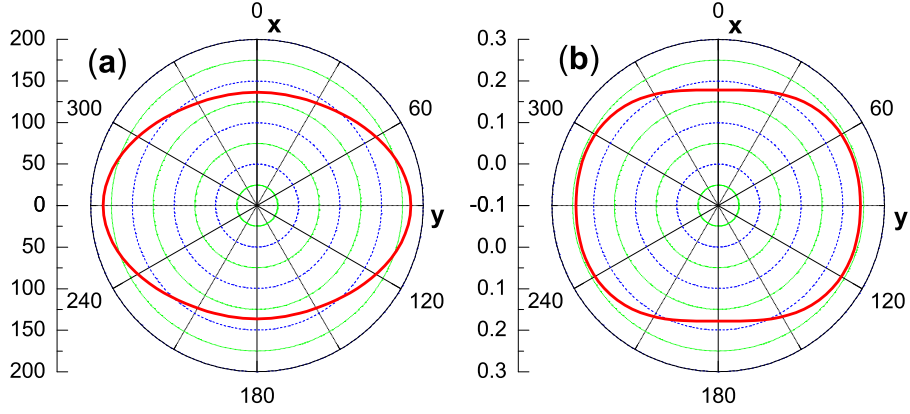}
\caption{(Color online)The Young's modulus and Possion's ratio of CCB monolayer as a function of the angle $\theta$. }\label{t1-1-1}
\end{figure*}
\begin{figure*}
  \includegraphics[width=15cm]{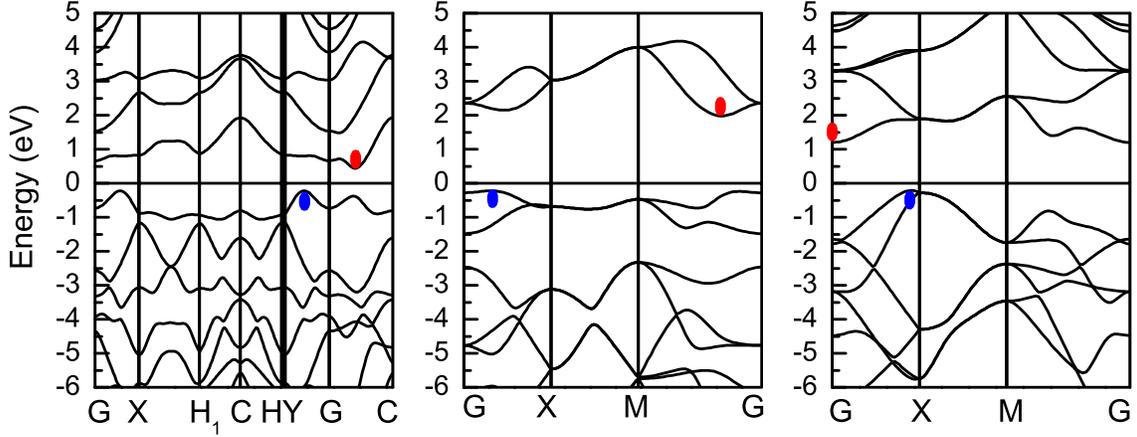}
  \caption{(Color online)The energy band structures  of  monolayer CCB (Left), CCC (Middle) and CBB (Right)  using GGA, and the VBM and CBM are marked by blue and red  ellipses, respectively.}\label{bi}
\end{figure*}

Compared with tetragonal structure of CCC and CBB,  the crystal structure of CCB changes into orthorhombic, and the optimized lattice constants $a$=$b$=3.771  $\mathrm{\AA}$ and $\gamma$=92.107. To confirm the dynamic stability of CCB monolayer, the
phonon spectrum  are calculated by VASP+Phonopy code with a supercell
of 4$\times$4$\times$1 using the finite displacement method\cite{pv5}. \autoref{t1-1} shows
the calculated phonon spectrum.  Although there are some negligibly small imaginary
frequencies near the G point due to calculation error, no imaginary
frequencies in the other q points throughout the Brillouin zone are observed, which  implies
the dynamic stability of CCB monolayer. The mechanical
stability of CCB monolayer can be examined by elastic
constants $C_{ij}$.
The  calculated  $C_{11}$=142.36  $\mathrm{Nm^{-1}}$,    $C_{22}$=193.07 $\mathrm{Nm^{-1}}$, $C_{12}$=34.40 $\mathrm{Nm^{-1}}$ and $C_{66}$=61.71 $\mathrm{Nm^{-1}}$,  which satisfy the  Born  criteria of mechanical stability\cite{ela1,ela}.
 The Young's modulus $C_{2D}(\theta)$ and Poisson's ratio $\nu (\theta)$  as a function of in-plane  $\theta$ can be attained on the basis of the elastic constants, as follows\cite{ela1}:
\begin{equation}\label{c2d}
C_{2D}(\theta)=\frac{C_{11}C_{22}-C_{12}^2}{C_{11}sin^4\theta+Asin^2\theta cos^2\theta+C_{22}cos^4\theta}
\end{equation}
\begin{equation}\label{c2d}
\nu (\theta)=\frac{C_{12}sin^4\theta-Bsin^2\theta cos^2\theta+C_{12}cos^4\theta}{C_{11}sin^4\theta+Asin^2\theta cos^2\theta+C_{22}cos^4\theta}
\end{equation}
In which  $A=(C_{11}C_{22}-C_{12}^2)/C_{66}-2C_{12}$ and $B=C_{11}+C_{22}-(C_{11}C_{22}-C_{12}^2)/C_{66}$. We  show the calculated $C_{2D}(\theta)$  and $\nu (\theta)$  in \autoref{t1-1-1}.
Both  the Young's modulus  $C_{2D}(\theta)$ and Poisson's ratio
$\nu (\theta)$ show  mechanical anisotropy. According to calculated $C_{2D}(\theta)$,  CCB monolayer is softer along the x than y  direction, which is due to B-B
bond along x direction and C-C bond along y direction. A high Young's modulus implies that the material is rigid, and the calculated results show  that strain can easily tune it's physical properties along x direction.  It is found that the  Poisson's ratio $\nu$ along x direction (0.179) is smaller than one along y direction (0.242).

\begin{table*}
\centering \caption{For CCB monolayer, elastic modulus ($C_{2D}$), effective mass ($m^*$), deformation potential ($E_l$), carrier mobility ($\mu_{2D}$)   and relaxation time ($\tau$) at 300 K.}\label{tab3}
  \begin{tabular*}{0.96\textwidth}{@{\extracolsep{\fill}}ccccccc}
  \hline\hline
Carrier type&    &$C_{2D}$ ($\mathrm{Nm^{-1}}$) & $m^*$ & $E_l$ (eV)& $\mu_{2D}$ ($\mathrm{cm^2V^{-1}s^{-1}}$)&$\tau$ (s)\\\hline\hline
Electrons   & x&  136.23&       1.57&       5.00         & 91.92&  $8.18\times10^{-14}$    \\
            &y&   184.76&       0.42 &      -1.15          & 8865.23   & $2.10\times10^{-12}$                                             \\
Holes   & x&       136.23&     -0.45&     -9.20          &114.63     & $2.90\times10^{-14}$                                            \\
             &y&   184.76&    -1.02&     6.97         &118.75      & $7.46\times10^{-14}$                             \\\hline\hline
\end{tabular*}
\end{table*}
\begin{figure*}
  \includegraphics[width=12cm]{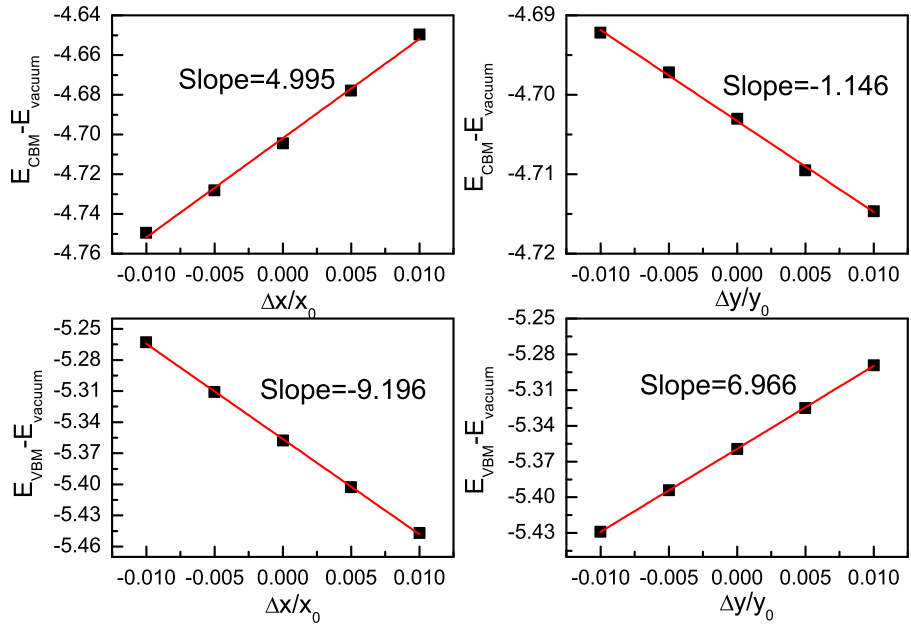}
  \caption{(Color online)With respect to the vacuum energy, the band energies of the VBM and CBM of CCB monolayer  as a function of lattice dilation along both x and y directions. The red solid lines are linear fitting curves, and the fitted slopes are given,  corresponding to the DP.}\label{t4}
\end{figure*}

A 2D material with piezoelectricity  not only  should   break inversion symmetry,  but also has a band gap.
The energy band structures of CCB along with CCC and CBB are plotted in \autoref{bi}.
It is clearly seen that CCB
is an indirect band-gap semiconductor with a band gap of 0.66 eV, which is smaller than  indirect gap  2.20 eV of CCC or   1.41 eV of CBB.
The valance band maximum (VBM)
lies on the G-Y path, while the conduction band
minimum (CBM) is located on the G-C path. In fact, the valance band extrema (VBE) along G-X path is very close to VBM due to $\gamma$ being very close to 90, and the difference is less than 1 meV. The carrier mobility of a semiconductors is  an important factor for  the application of electron device.
The carrier mobility of a 2D material  ($\mu_{2D}$) by the deformation potential (DP) theory proposed by Bardeen and Shockley\cite{dp} is defined as:
\begin{equation}\label{u2d}
  \mu_{2D}=\frac{e\hbar^3C_{2D}}{K_BTm^*m_dE_l^2}
\end{equation}
where $T$, $m^*$ and $m_d$ ($\sqrt{m_xm_y}$) are the  temperature, the effective mass in the transport direction and   the average effective mass.
The $C_{2D}$ is the Young's modulus derived from  elastic
constants $C_{ij}$. In addition, $E_l$ is the DP constant defined by $E_l=\Delta E/\delta$  ($\delta=\Delta l/l_0$)
, where $\Delta E$ is the energy shift of the band edge of CBM or VBM with respect
to the vacuum level  after applying uniaxial strain. After attaining $\mu_{2D}$,  the  relaxation
time $\tau$ can be attained by:
\begin{equation}\label{t}
    \tau=\mu_{2D}m^*/e
\end{equation}

According to  DP theory,  the rectangular supercell is used to calculate  carrier mobilities of monolayer CCB along the  x and y directions in \autoref{t0-1}.
The calculated effective masses for electrons (CBM) and holes (VBM) with GGA  are shown in \autoref{tab3}.
The band energies of the VBM and CBM  with respect to the vacuum energy as a function of $\Delta x/x$ and $\Delta y/y$ are plotted in \autoref{t4}, and
the DP constant $E_l$ is calculated  by linearly fitting these data.
The carrier mobility and relaxation
time  for the electrons and holes of monolayer CCB on the basis of the calculated effective mass, elastic
constant, and deformation potential constant   are calculated along x and y directions at 300 K, which are summarized in \autoref{tab3}.
It is found that the electron mobility along y direction
(8865.23 $\mathrm{cm^2V^{-1}s^{-1}}$) is almost 97 times larger than that along
x direction (91.92 $\mathrm{cm^2V^{-1}s^{-1}}$). However, the hole mobilities along both x (114.63 $\mathrm{cm^2V^{-1}s^{-1}}$) and y (118.75 $\mathrm{cm^2V^{-1}s^{-1}}$) directions are very low, which is close to electron mobility along x direction.

In fact, the  CCB is a representative Janus monomlayer. The asymmetric along the z direction  with respect to central C atomic layer   results
in an out-of-plane piezoelectricity.  However,  mirror symmetry  along x and y directions  gives rise to disappeared in-plane piezoelectricity.
This means that an in-plane stress or strain  can only  induce a polarization change vertical to the plane.
Based on calculated $C_{ij}$ and $e_{ij}$, the  out-of-plane piezoelectric strain coefficient $d_{31}$ and $d_{32}$ can be attained, and the  corresponding values are   -0.505 pm/V  and 0.273 pm/V, respectively.
The out-of-plane  $d_{31}$  is obviously
higher than out-of-plane ones of  many other 2D  materials, such as
functionalized h-BN (0.13 pm/V)\cite{y1}, MoSSe (0.29 pm/V)\cite{q7-2}, SbTeI (-0.37 pm/V)\cite{q7-2-1}, MoSTe (0.4 pm/V)\cite{q7-1}, Janus group-III materials (0.46 pm/V)\cite{y2} and $\alpha$-$\mathrm{In_2Se_3}$ (0.415 pm/V)\cite{y3}.  It is lower than one of BiTeI (-0.66 pm/V)\cite{q7-2-1}, $\mathrm{Sc_2CO_2}$ (0.78 pm/V)\cite{y4} or $\mathrm{La_2CO_2}$ (0.65 pm/V)\cite{y4}. However,  the  monolayer BiTeI, $\mathrm{Sc_2CO_2}$  or $\mathrm{La_2CO_2}$ not only has out-of-plane piezoelectricity, but also has in-plane piezoelectricity.
 Some related data are summarized in \autoref{tab-y}.
Although the $d_{36}$  (CCC monolayer) and $d_{31}$ or $d_{32}$ (CCB monolayer) are all
related to the out-of-plane piezoelectric effects, they are obviously different.
The $d_{36}$ represents the piezoelectric response between the out-of-plane polarization
and the in-plane shearing deformation,  while $d_{31}$ or $d_{32}$ is referred to as the
coupling between the out-of-plane polarization and the in-plane normal strain.

\section{Conclusion}
In summary, the related piezoelectric effects of  CCC monolayer are studied by using reliable first-principles calculations.
Due to special symmetry for CCC monolayer, only  out-of-plane $d_{36}$  exists, but it is very small.
Firstly, we use both  biaxial and uniaxial  strains to tune  piezoelectric properties of CCC monolayer.
At -2\%  biaxial (-4\% uniaxial) strain, the $d_{36}$ is  increased only by 3.1\% (13.9\%), and the enhancement is very small.
 Inspiring from the already synthesized MoSSe monolayer,  a Janus CCB monolayer with dynamic and mechanical stability  is  constructed, and the asymmetric along the z direction and   mirror symmetry  along x and y directions induce pure  out-of-plane piezoelectricity. It is also  found that CCB monolayer is a semiconductor, which is necessary for  piezoelectric application. The calculated  $d_{31}$ and $d_{32}$  are   -0.505 pm/V  and 0.273 pm/V, respectively, and the out-of-plane piezoelectric effect are obviously improved, compared with  $d_{36}$ of CCC monolayer.
 The out-of-plane piezoelectricity $d_{31}$ of CCB monolayer is  higher compared with other many 2D known materials.
 The very high  room-temperature electronic mobility (8865.23 $\mathrm{cm^2V^{-1}s^{-1}}$) along y direction  is predicted, which is very higher than that of Si (about 1400 $\mathrm{cm^2V^{-1}s^{-1}}$). Our works not
only supply an experimental proposal for achieving  large pure out-of-plane piezoelectric effect
, but also offer an insight into piezoelectric effect of  penta-graphene.

\begin{acknowledgments}
This work is supported by the Natural Science Foundation of Shaanxi Provincial Department of Education (19JK0809). We are grateful to the Advanced Analysis and Computation Center of China University of Mining and Technology (CUMT) for the award of CPU hours and WIEN2k/VASP software to accomplish this work.
\end{acknowledgments}


\begin{references}
\bibitem{w1}S. Zhang, J. Zhou, Q. Wang, X. Chen, Y. Kawazoe and  P. Jena,  Proc. Natl. Acad. Sci. Unit. States Am. \textbf{112}, 2372 (2015).

\bibitem{w2}F. Li, K. Tu, H. Zhang and  Z. Chen,  Phys. Chem. Chem. Phys.  \textbf{17}, 24151 (2015).

\bibitem{w3} Y. Ma, L. Kou, X. Li, Y. Dai and  T. Heine,  NPG Asia Mater.  \textbf{8}, e264 (2016).



\bibitem{w4} J. Li, X. Fan, Y. Wei, H. Liu, S. Li, P. Zhao and  G. Chen, Sci. Rep.  \textbf{6}, 33060 (2016).


\bibitem{w5}H. Sun, S. Mukherjee and  C. V. Singh,  Phys. Chem. Chem. Phys. \textbf{18}, 26736 (2016).


\bibitem{w6}F. Q. Wang, J. Yu, Q. Wang, Y. Kawazoe and  P. Jena,  Carbon \textbf{105}, 424 (2016).


\bibitem{w6-1}W. Xu, G. Zhang and B. Li, J. Chem. Phys. \textbf{143}, 154703 (2015).

\bibitem{w6-2}J. Sun, Y. G. Guo, Q. Wang and  Y.  Kawazoe, Carbon \textbf{145}, 445 (2019).

\bibitem{w7}L. L. Liu, Y. Wang, C. P. Chen, H. X. Yu, L. S. Zhao and  X. C. Wang,  RSC Adv. \textbf{7}, 40200 (2017).


\bibitem{q4}W. Wu and Z. L. Wang, Nat. Rev. Mater. \textbf{1}, 16031 (2016).



\bibitem{q5} W. Wu, L. Wang, Y. Li, F. Zhang, L. Lin, S. Niu, D. Chenet,
X. Zhang, Y. Hao, T. F. Heinz, J. Hone and Z. L. Wang,
Nature \textbf{514}, 470 (2014).


\bibitem{q6}H. Zhu, Y. Wang, J. Xiao, M. Liu, S. Xiong, Z. J. Wong, Z. Ye,
Y. Ye, X. Yin and X. Zhang, Nat. Nanotechnol. \textbf{10},
151 (2015).

\bibitem{q8}A. Y. Lu, H. Zhu, J. Xiao, C. P. Chuu, Y. Han, M. H. Chiu,
C. C. Cheng, C. W. Yang, K. H. Wei, Y. Yang, Y. Wang,
D. Sokaras, D. Nordlund, P. Yang, D. A. Muller, M. Y. Chou,
X. Zhang and L. J. Li, Nat. Nanotechnol. \textbf{12}, 744 (2017).

\bibitem{q7}L. Dong, J. Lou and V. B. Shenoy, ACS Nano, \textbf{11},
8242 (2017).


\bibitem{q9}M. N. Blonsky, H. L. Zhuang, A. K. Singh and R.  G. Hennig,  ACS Nano, \textbf{9},
9885 (2015).

\bibitem{q10}R. X. Fei, We. B. Li, J. Li and L. Yang, Appl. Phys. Lett.  \textbf{107}, 173104 (2015)


\bibitem{q11}K. N. Duerloo, M. T. Ong and E. J. Reed, J. Phys. Chem. Lett. \textbf{3}, 2871 (2012).


\bibitem{q12}Y. Chen,  J. Y. Liu,  J. B. Yu,  Y. G. Guo and Q. Sun, Phys. Chem. Chem. Phys.
 \textbf{21}, 1207 (2019).


 \bibitem{pv6}X. Wu, D. Vanderbilt and  D. R.  Hamann, Phys. Rev. B  \textbf{72}, 035105 (2005).


\bibitem{1}P. Hohenberg and W. Kohn, Phys. Rev. \textbf{136},
B864 (1964); W. Kohn and L. J. Sham, Phys. Rev. \textbf{140},
A1133 (1965).

\bibitem{pv1} G. Kresse, J. Non-Cryst. Solids \textbf{193}, 222 (1995).

\bibitem{pv2} G. Kresse and J. Furthm$\ddot{u}$ller, Comput. Mater. Sci. 6, \textbf{15} (1996).

\bibitem{pv3} G. Kresse and D. Joubert, Phys. Rev. B \textbf{59}, 1758 (1999).

\bibitem{pbe}J. P. Perdew, K. Burke and M. Ernzerhof, Phys. Rev. Lett. \textbf{77}, 3865 (1996).









\bibitem{r1}N. Jena, Dimple, S. D.  Behere  and A. D. Sarkar, J. Phys. Chem. C  \textbf{121}, 9181 (2017).

\bibitem{r2}S. D. Guo, X. S. Guo, Y. Y. Zhang and K. Luo, arXiv:1910.08700 (2019).

\bibitem{r3}Dimple, N. Jena, A. Rawat, R.  Ahammed,
M. K. Mohanta and A. D. Sarkar, J. Mater. Chem. A  \textbf{6},
24885 (2018).




\bibitem{ela1}E. Cadelano, P. L. Palla, S. Giordano and L. Colombo,  Phys. Rev. B  \textbf{82}, 235414 (2010).
\bibitem{ela}R. C. Andrew, R. E. Mapasha, A. M. Ukpong and N. Chetty, Phys. Rev. B \textbf{85}, 125428 (2012).

\bibitem{pv5}A. Togo, F. Oba, and I. Tanaka, Phys. Rev. B \textbf{78}, 134106
(2008).

\bibitem{dp}S. Bruzzone and G. Fiori, Appl. Phys. Lett. \textbf{99}, 222108 (2011).

\bibitem{y1}A. A. M. Noor, H. J. Kim and  Y. H. Shin, Phys. Chem.
Chem. Phys. \textbf{16}, 6575 (2014).



\bibitem{q7-2}S. D. Guo, X. S. Guo, R. Y. Han and Y. Deng,  Phys. Chem. Chem. Phys. (2019). DOI: 10.1039/C9CP04590B.
\bibitem{q7-2-1}S. D. Guo, X. S. Guo, Z. Y.  Liu, Y. N. Quan,	arXiv:1909.13227 (2019).


\bibitem{q7-1}M. Yagmurcukardes, C. Sevik and F. M. Peeters, Phys. Rev. B  \textbf{100}, 045415 (2019).




\bibitem{y2}Y. Guo, S. Zhou, Y. Z. Bai, J. J. Zhao,  Appl. Phys. Lett. \textbf{110},  163102 (2017).


\bibitem{y3}L. Hu and X. R. Huang, RSC Adv. \textbf{7},  55034 (2017).


\bibitem{y4}J. Tan, Y. H. Wang, Z. T. Wang et al., Nano Energy
\textbf{65}, 104058 (2019).


\end{references}
\end{document}